\documentclass[fleqn,twoside,twocolumn,nofootinbib,showkeys]{revtex4} 
\usepackage[sec,nocpr]{ujp} 

\begin{document}
\title[Rotation of Plasma Layers with Various
Densities]%
{ROTATION OF PLASMA LAYERS WITH VARIOUS DENSITIES IN CROSSED
$\mathbf{E}\times\mathbf{B}$ FIELDS}%
\author{Yu.V.~Kovtun}%
\affiliation{National Science Center Kharkiv Institute of Physics
and Technology,\\
Nat. Acad. of Sci. of Ukraine}%
\address{1, Akademichna Str., Kharkiv 61108, Ukraine}%
\email{Ykovtun@kipt.kharkov.ua}
\author{E.I.~Skibenko}%
\affiliation{National Science Center Kharkiv Institute of Physics
and Technology,\\
Nat. Acad. of Sci. of Ukraine}%
\address{1, Akademichna Str., Kharkiv 61108, Ukraine}%
\author{A.I.~Skibenko}
\affiliation{National Science Center Kharkiv Institute of Physics
and Technology,\\
Nat. Acad. of Sci. of Ukraine}%
\address{1, Akademichna Str., Kharkiv 61108, Ukraine}%
\email{Ykovtun@kipt.kharkov.ua}%
\author{V.B.~Yuferov}%
\affiliation{National Science Center Kharkiv Institute of Physics
and Technology,\\
Nat. Acad. of Sci. of Ukraine}%
\address{1, Akademichna Str., Kharkiv 61108, Ukraine}%
\email{Ykovtun@kipt.kharkov.ua}%
\udk{533.915} \pacs{52.80.Sm; 52.65.Kj}\razd{\secv}

\autorcol{Yu.V.\hspace*{0.7mm}Kovtun, E.I.\hspace*{0.7mm}Skibenko,
A.I.\hspace*{0.7mm}Skibenko et al.}

\setcounter{page}{450}%

\begin{abstract}
The rotational velocity of plasma layers with various densities in
a pulsed reflex-discharge plasma is studied with the use of
the two-frequency microwave fluctuation reflectometry. The difference
between the angular rotational velocities of plasma layers with
different densities is revealed, and their time dependences are
determined. The rotational velocity of plasma layers is found to
increase with the magnetic field induction. On the basis
of the experimental data obtained, the radial electric field strength in
the plasma layers concerned is evaluated.
\end{abstract}
\keywords{plasma, fluctuation reflectometry, plasma layers.}
\maketitle

\section{Introduction}

Experimental researches of plasma in crossed
$\mathbf{E}\times\mathbf{B}$ fields were started at the beginning of
the 1950s in the framework of the activity aimed at the controlled
thermonuclear fusion \cite{1,2,3}, although the works dealing with
gas discharges in crossed fields were initiated at the end of the
19th and the beginning of the 20th centuries \cite{4,5,6}. Nowadays,
the experiments devoted to various
physical and applied aspects concerning the plasma in crossed fields are intensively carried
out on various experimental installations, such as MCX \cite{7},
MISTRAL \cite{8}, MBX \cite{9}, ALEXIS \cite{10}, and a number of
others. One of the features for plasma created and held in crossed
$\mathbf{E}\times\mathbf{B}$ fields is its drift rotation. Under
certain conditions, various instabilities can develop in the
rotating plasma, which may bring about, e.g., the heating of the
plasma ion component \cite{11,12}. In the case of multicomponent
plasma, the rotation of a plasma column gives rise to the spatial
separation of the ion component \cite{13}. There can be several variants
at that. First, plasma contains ions with identical masses, but in
different charged states. In this case, ions with a large $Z$ drift
toward the center of the plasma column. Second, plasma contains ions
with different masses. In this case, owing to centrifugal forces
that arise in the rotating plasma, the ions become radially separated
according to their masses. The third variant, in essence, is a
combination of the first and second ones. The efficiency of the radial
ion separation depends on the rotational velocity \cite{13}. In
connection with the aforesaid, the determination of the plasma
rotational velocity is of a certain interest. On the other hand, the
experimental research of the rotational velocity in plasma was carried
out mainly for the plasma of various gases or their mixtures
\cite{14,15}, as well as the gas-metal plasma \cite{16}. The study
of rotation in a gas-metal plasma has the pure physical
importance, as well as the applied significance.

Therefore, this work is aimed at studying the rotational velocity in
the multicomponent gas-metal plasma embedded in crossed
$\mathbf{E}\times\mathbf{B}$ fields. It continues our earlier
researches of a multicomponent gas-metal plasma created in the
pulsed reflex discharge \cite{17,18,19,20,21}.

\section{Plasma Motion in Crossed \boldmath$\mathbf{E}\times\mathbf{B}$
Fields}

In the non-relativistic case, the equation of motion for a charge in an
electromagnetic field looks like \cite{22,23}
\begin{equation}
m\frac{d\mathbf{v}}{dt}=q\mathbf{E}+q\left[ \mathbf{v\times
B}\right]\!, \label{1}
\end{equation}

\noindent where
$\mathbf{E}=-\frac{1}{c}\frac{d\mathbf{A}}{dt}-\bigtriangledown\phi$
is the electric field strength,
$\mathbf{B}=\mathrm{rot~}\mathbf{A}$ the magnetic field induction,
$q=Ze$ the ion charge,
$e$ the elementary charge, $m$ the particle's mass, $\mathbf{A}$ the
vector potential of the field, $\mathbf{v}$ the particle's velocity, and
$\phi$ the scalar potential of the field. The motion of a charged
particle in
crossed electric and magnetic fields, in particular, in an axially
symmetric
reflex discharge, is described by the following system of differential
equations (in the cylindrical coordinates):
\begin{equation}
\left\{\!
\begin{array}{lcl}
\displaystyle m \left(\ddot{r} - r\dot{\varphi}^2\right) = q E_r + q
r
\dot{\varphi} B_z,\\[1mm]
\displaystyle m \frac {1}{r} \frac {d}{dt} \left( r^2
\dot{\varphi}\right)=q
\left(\dot{z}B_r - \dot{r}B_z \right)\!,\\[3mm]
\displaystyle m \ddot{z} = q E_z - q r \dot{\varphi} B_r .
\end{array}
\right.\label{2}
\end{equation}

\noindent For the state equilibrated in the radial direction, the
first equation of system~(\ref{2}) looks like
\begin{equation}
mr\omega^{2}+qr\omega B_{z}+qE_{r}=0,\label{3}
\end{equation}
where $\dot{\varphi}=\omega$ is the angular velocity. As is seen
from Eq.~(\ref{3}), the particle is subjected to the action of the
following forces in the radial direction perpendicular to the
particle rotation axis: (i)~the centrifugal force associated with
particle's motion, $mr\omega ^{2}$; (ii)~the magnetic force
associated with the action of magnetic field on the particle,
$qr\omega B_{z}$; and (iii)~the electric force produced by the
electric field, $qE_{r}$. The drift of the particle is determined by
the relation \cite{1} \begin{equation*}
v_{i\varphi}=-\frac{E_{r}}{B_{z}}-\frac{m_{i}v_{i\varphi}^{2}}{rqB_{z}}
,
\end{equation*}
\noindent where $v_{i\varphi}=\omega r$ is the rotational velocity.
The
solution of the quadratic equation (\ref{3}) for ions with regard for
the
particle charge and the field direction (it is directed radially toward
the
rotation axis, so that the positive potential increases with the distance
to
the rotation axis) reads
\begin{equation}
v_{i\varphi} = \frac {\Omega_i r}{2} \left(\!\!-1 + \sqrt {1 + \frac
{4 E_r}{r B_z \Omega_i} } \right)\!,  \label{4}
\end{equation}

\noindent where $\Omega_{i}=\frac{qB_{z}}{m_{i}}$ is the Larmor
frequency of
ion rotation, $r$ the rotation radius, and $m_{i}$ the ionic mass. For
an
electron, the centrifugal force $mr\omega^{2}$ is small and, hence, can
be
neglected. Then, the rotational velocity equals
\begin{equation}
v_{e\varphi}=\frac{E_{r}}{B_{z}}.   \label{5}
\end{equation}

\begin{figure}
\vskip1mm
\includegraphics[width=6cm]{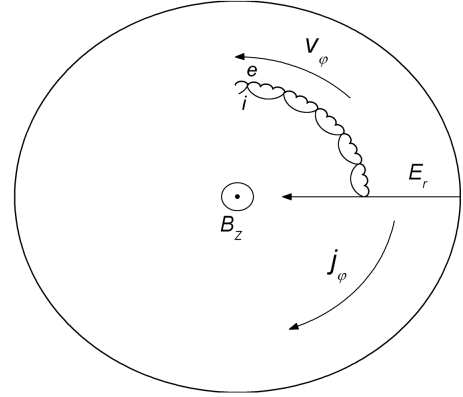}
\vskip-3mm\caption{Rotation of electrons $e$ and ions $i$ in crossed
$\mathbf{E}\times\mathbf{B} $ fields  }
\end{figure}

\noindent Owing to the difference between the rotational velocities
$v_{\varphi}$ for electrons and ions, there emerges a circular electric
current, with its density being proportional to the difference between
the
corresponding values for both plasma components \cite{1}.
\begin{equation}
j_{\varphi} = N q \left(v_{i\varphi} - v_{e\varphi}
\right)\!.\label{6}
\end{equation}

\noindent The current induces the diamagnetic effect and diminishes
the magnetic field induction in plasma, in addition to a reduction
associated with the usual plasma diamagnetism. As an illustration of
the aforesaid, Fig.~1 exhibits the trajectories of particles and the
direction of their rotation in the crossed
\mbox{$\mathbf{E}\times\mathbf{B}$ fields.}

In the framework of the macroscopic approach, the average velocity of all
particles (of a certain kind) included into an element of
volume
is considered. Then, every plasma component can be regarded as a fluid,
the
motion of which is described by the macroscopic velocity. In the
two-fluid
magnetohydrodynamic (MHD) model, the equations of motion look like
\cite{24}
\[ N_e m_e \left(\!\frac {d \mathbf {v}_e}{dt} + \left(\mathbf {v}_e \nabla \right) \mathbf {v}_e \!\right) = \]\vspace*{-7mm}
\begin{equation}
= N_e q \left[\mathbf {E} +  \mathbf {v}_e \times  \mathbf
{B}\right]- \nabla P_e + R,\label{7}
\end{equation}\vspace*{-7mm}
\[ N_i m_i \left(\!\frac {d \mathbf {v}_i}{dt} + \left(\mathbf {v}_i \nabla\right) \mathbf {v}_i\!\right) = \]\vspace*{-7mm}
\begin{equation}
= N_i q \left[\mathbf {E} +  \mathbf {v}_i \times  \mathbf
{B}\right]- \nabla P_i - R  ,\label{8}
\end{equation}

\begin{figure}
\vskip1mm
\includegraphics[width=\column]{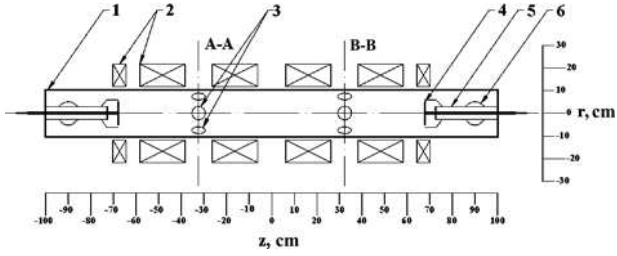}
\vskip-3mm\caption{Diagram of the pulsed reflex-discharge
installation: (\textit{1})~discharge chamber (anode),
(\textit{2})~magnetic system, (\textit{3})~diagnostic ports,
(\textit{4})~cathodes, (\textit{5})~insulator, (\textit{6})~vacuum
pumping system, (A--A and B--B)~cross-sections of the diagnostic
port arrangement  }
\end{figure}

\begin{table}[b]
\noindent\caption{}\vskip3mm\tabcolsep9.2pt
\noindent{\footnotesize\begin{tabular}{|c| l| c| }
 \hline \multicolumn{1}{|c}
{\rule{0pt}{5mm}No.} & \multicolumn{1}{|c}{Parameter}&
\multicolumn{1}{|c|}{Value}\\[2mm]%
\hline%
\rule{0pt}{5mm}\,\,1 &   Discharge voltage, kV & $\le $4.5 \\%
\,\,2 & Discharge current, kA & $\le $2 \\%
\,\,3 &   Current pulse duration, ms & $\le $1\\%
\,\,4 &   Capacitor bank, $\mu\mathrm{F}$ & 560 \\%
\,\,5 &    Energy duty, kJ& $\le 6$\\%
\,\,6 &    Magnetic field induction, T & $\le $0.9 \\%
\,\,7 & Mirror ratio $R$ & 1.25\\%
\,\,8 & Vacuum chamber volume, $~\mathrm{cm}^{3}$ &  $\sim $6.6~$
\times 10^4$
\\%
\,\,9 & Initial pressure, Pa & $1.33 \times 10^{-4}$ \\%
10 &    Working pressure, Pa & 0.133--4.7\\%
11 & Diameter of cathodes, cm &10 \\%
12 &   Cathode material & Ti\\%
13 &   Ignition gas  & Ar \\%
14 & Plasma volume, $~\mathrm{cm}^{3}$ & $\sim $10$^4$ \\[2mm]%
\hline
\end{tabular}}
\end{table}

\noindent where  $N_{e}$ and $N_{i}$ are the concentrations of
electrons and ions, respectively; $m_{e}$ and $m_{i}$ are the masses
of electron and ion, respectively; $\mathbf{v}_{e}$ and
$\mathbf{v}_{i}$ are the velocities of electrons and ions,
respectively; $P_{e}$ nad $P_{i}$ are the pressures of the electron
and ion components, respectively; $R=-N_{e}m_{e}\left(
\mathbf{v}_{e}-\mathbf{v}_{i}\right) v_{ei}$ is the friction force;
and $v_{ei}$ is the frequency of collisions between the electrons
and ions. For the state equilibrated in the radial direction, from
Eq.~(\ref{8}) and assuming $v_{ir}=v_{er}=0$, we obtain that, for
ions,
\begin{equation}
m_{i}r\omega^{2}+qr\omega B_{z}+qE_{r}-\frac{1}{N_{i}}\nabla
P_{i}=0. \label{9}
\end{equation}

\noindent For  electrons, a similar equation can be obtained.
Equation (\ref{9}) transforms to the form
\begin{equation}
v_{i\varphi}=-\frac{E_{r}}{B_{z}}-\frac{m_{i}v_{i\varphi}^{2}}{rqB_{z}}
+ \frac{1}{N_{i}qB_{z}}\bigtriangledown P_{i}. \label{10}
\end{equation}

\noindent Whence,  one can see that the rotational velocity consists
of the drift in the electric field, the centrifugal drift, and the
diamagnetic drift. The solution of quadratic Eq.~(\ref{9}) with
regard for the charge of an ion and the field direction looks like
\begin{equation}
v_{i\varphi} = \frac {\Omega_i r}{2} \left(\!\!-1 + \sqrt {1 + \frac
{4 E_r}{r B_z \Omega_i} + \frac {4 \nabla P_i}{r m_i N_i
\Omega_{i}^2} } \right)\! . \label{11}
\end{equation}

\noindent For electrons, when the centrifugal force is neglected, we
obtain
\begin{equation}
v_{e\varphi}=\frac{E_{r}}{B_{z}}-\frac{1}{N_{e}qB_{z}}\bigtriangledown
P_{e}. \label{12}
\end{equation}
If the diamagnetic drift is not taken into account, Eqs.~(\ref{11}) and
(\ref{12}) look similarly to Eqs.~(\ref{4}) and (\ref{5}), respectively.

In the one-fluid MHD model, the equation of motion for the plasma as a whole
reads
\begin{equation}
\rho \left(\!\frac {d \mathbf {v}}{dt} + \left(\mathbf {v} \nabla
\right) \mathbf {v}\!\right) =  \mathbf {j} \times \mathbf {B}-
\nabla P , \label{13}
\end{equation}

\noindent where $P=P_{e}+P_{i}$ is the total plasma
pressure;~$\rho$~is the plasma mass density ($\rho\approx
N_{p}m_{i}$, because $m_{e}\ll$ $\ll m_{i}$);
$\mathbf{v}=N_{p}\left(
m_{i}\mathbf{v}_{i}+m_{e}\mathbf{v}_{e}\right) /\rho$ is the
velocity of the plasma charged component; and
$\mathbf{j}=eN_{p}\times$ $\times\left(
\mathbf{v}_{i}-\mathbf{v}_{e}\right) $ is the current density
($N_{i}=N_{e}=N_{p}$). Decomposing Eq.~(\ref{13}), we obtain
\begin{equation}
\rho\left(\!
\frac{dv_{r}}{dt}+v_{r}\frac{dv_{r}}{dr}-\frac{dv_{\varphi}^{2}}{r}\!\right
) =j_{\varphi}B_{z}-\frac{dP}{dr},  \label{14}
\end{equation}\vspace*{-7mm}
\begin{equation}
\rho\left(\!\frac {d v_\varphi}{dt} + v_r \frac {d v_\varphi}{dr}+
\frac { v_{\varphi} v_r}{r}\! \right) = -j_r B_z  .\label{15}
\end{equation}

\section{Experimental Installation\\ and Diagnostic Techniques}

The rotational velocity of the gas-metal plasma created in a
high-current pulsed reflex discharge was experimentally studied on
an installation exhibited in Fig.~2. The gas-metal plasma was formed
as a result of a discharge in an environment consisting of a working
gas, Ar, and a sputtered material of cathodes. The cathodes were
fabricated from monometallic Ti. The supply of the cathode material
(Ti) into plasma was confirmed by spectrometric measurements
\cite{17,18,19}, with the amount of titanium in plasma could be at a
level of 40--50\% \cite{20}. The maximum plasma density amounted to
$N_{p}\sim10^{14}$~\textrm{cm}$^{-3}$. The main electrophysical
parameters of the installation and the discharge are quoted in
Table. A pulsed magnetic field of the mirror configuration and 18~ms
in duration was created by a solenoid composed of six coils (see
Fig.~2). The time dependences of the magnetic field induction for
various voltages $U_{B}$ at a capacitor bank of the magnetic system
are depicted in Fig.~3.

A variety of diagnostic techniques for the determination of a plasma
rotational velocity are known. These are the corpuscular-optical
(the char\-ge-ex\-chan\-ge spectroscopy), optical (the Doppler
spectrometry), microwave (the Doppler and fluctuation
reflectometry), and probe methods. The choice of that or another
diagnostic technique is mainly associated with a possibility of its
application under specific experimental conditions. To study the
rotation in a multicomponent gas-metal plasma, the most suitable are
contactless methods including the optical and microwave methods.
Both methods have their specific advantages and shortcomings. The
main advantage of the optical method consists in the capability of
measuring the rotational velocity of different plasma ions, although
rather a complicated hardware is required for this purpose~--
especially if the radial dependence of the rotational velocity is
determined. Microwave methods are simple enough from this aspect;
however, they allow the density and rotational velocity of only the
electron plasma component to be determined. In this connection, we
chose the method of microwave fluctuation reflectometry. This method
is based on the determination of the auto- and cross-correlation
functions for two poloidally separated microwave signals reflected
from a plasma layer with the same density. The auto- and
cross-correlation functions can be calculated by the formulas
\begin{equation}
c_{xx}\left( \tau_{k}\right) =\frac{1}{N}\sum_{t=0}^{N-1}x\left(
t\right) x\left( t+\tau_{k}\right)\!,   \label{16}
\end{equation}\vspace*{-5mm}
\begin{equation}
c_{xy}\left( \tau_{k}\right) =\frac{1}{N}\sum_{t=0}^{N-1}x\left(
t\right) y\left( t+\tau_{k}\right)\!,   \label{17}
\end{equation}

\begin{figure}
\vskip1mm
\includegraphics[width=7cm]{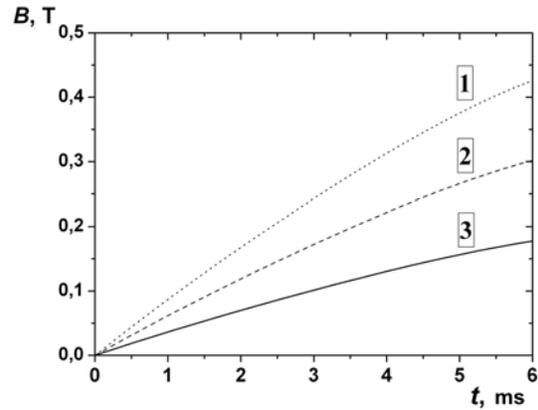}
\vskip-3mm\caption{Time dependence of the magnetic field induction.
$U_{B}=2$ (\textit{1}), 1.4 (\textit{2}), and 0.84$~\mathrm{kV}$
(\textit{3})  }
\end{figure}

\noindent where $c_{xx}\left( \tau_{k}\right) $ and $c_{xy}\left(
\tau _{k}\right) $ are the auto- and cross-correlation,
respectively, functions for the signals $x(t)$ and $y(t)$; $N$ is
the number of pixels in the $x(t)$- and $y(t)$-realizations; and
$\tau_{k}$ is the time lag between two signals. In contrast to the
measurements of the Doppler frequency shift for a microwave signal,
when the sounding direction is inclined and the reflection points do
not coincide with the layer where $N_{p}=$ $=N_{\rm cr.}$, the
correlation method is based on the normal sounding. Therefore, the
spatial position of the layer and its rotational velocity can be
determined simultaneously, which is especially important for
diagnosing the pulsed discharge plasma. In the case of a profile
with the circular symmetry, the velocity of poloidal plasma rotation
is found from the relation
\begin{equation}
v_{\varphi}=\frac{\Delta l}{\Delta t}=\frac{\Delta\varphi r}{\Delta
t},
\label{18}
\end{equation}

\noindent where $\Delta\varphi$ is the angular distance between
the points of the reflected wave detection, $r$ is the position of
the reflecting layer determined from the phase shift of a
reflected wave; and $\Delta t$ is either the shift of the
cross-correlation function (CCF) maximum in time or the period of
the auto-correlation function (ACF).

One of the features of this work consists in the application of
two-frequency microwave fluctuation reflectometry for the
determination of the rotational velocity of plasma layers with
$N_{p}=N_{\rm cr.}^{1,2}$ in the reflex discharge. The plasma
sounding frequency was so selected that, firstly, a layer with
$N_{p}= N_{\rm cr.}^{1,2}$ should exist in the generated plasma,
and, secondly, plasma layers with different $N_{\rm cr.}^{1,2}$
should be separated from each other by a distance that is larger
than the uncertainty in the position of the reflecting layer
\cite{25}. Therefore, two sounding frequencies were selected,
$f_{1}=37.13~\mathrm{GHz}$ and $f_{2}=72.88~\mathrm{GHz,}$ which
corresponded to $N_{\rm cr.}^{1}=1.7\times 10^{13}~\mathrm{cm}^{-2}$
and $N_{\rm cr.}^{2}=6.5\times10^{13}$~\textrm{cm}$^{-3}$,
respectively. %
\begin{figure}
\vskip1mm
\includegraphics[width=5cm]{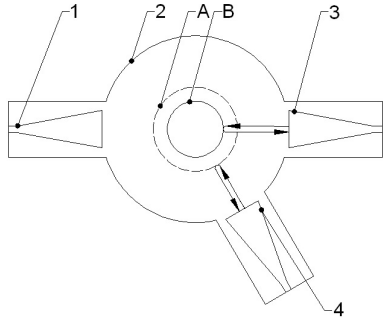}
\vskip-3mm\caption{ Diagram of the arrangement of antennas for
correlation and interferometric measurements: ({\it 1})~receiving
antenna of interferometer, ({\it 2})~vacuum chamber, ({\it 3} and
{\it 4}) transmit-receive antennas of a reflectometer. ${\bf A}$ and
${\bf B}$ are plasma layers with $N_{p}=N_{\rm cr.}^{1,2}$
}\vskip5mm
\end{figure}%
\begin{figure}
\includegraphics[width=6.7cm]{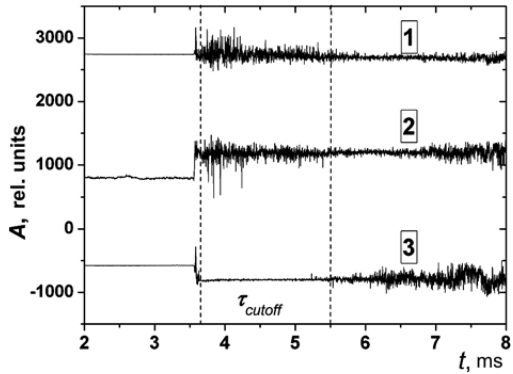}
\vskip-3mm\caption{Oscillograms of signals reflected from plasma
layers with $N_{p}=N_{\rm
cr.}^{1}=1.7\times10^{13}$~\textrm{cm}$^{-3}$ (\textit{1} and
\textit{2}) and the signal from a UHF-interferometer, $\lambda=8$~mm
(\textit{3})  }\vskip5mm
\end{figure}%
\begin{figure}[h!]
\includegraphics[width=7.5cm]{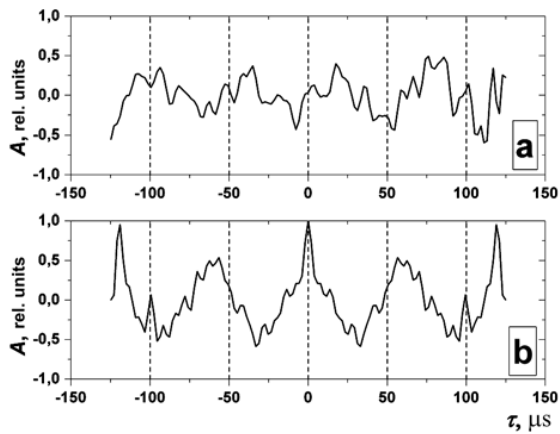}
\vskip-3mm\caption{Cross-correlated ({\it a}) and auto-correlated
({\it b}) functions for a reflected signal  }\vskip-4mm
\end{figure}%
The plasma was sounded using an ordinary (O) wave across the plasma
column and in the same cross-section for both frequencies. The
arrangement diagram for microwave antennas is shown in Fig.~4. The
application of two combined antennas poloidally separated by an
angle of 60$^{\circ}$ enabled the ACF and CCF to be used depending
on the scheme of antenna connection; i.e. each of the antennas was
used only for one sounding frequency, or both antennas were
simultaneously supplied with microwave signals of different
frequencies. Simultaneously with reflectometric measurements, the
maximum, $N_{p}=N_{\rm cr.}^{1,2}$, and average densities were
measured with the help of an UHF interferometer. It allowed us to
determine the lifetime interval for the layer with the critical
density. The signals were registered with the use of an
analog-to-digital converter with a frequency of
20~MHz.\vspace*{-2mm}

\section{Experimental Results}

The rotational velocity in the gas-metal plasma Ar~+~Ti was
experimentally studied under the following initial conditions for
the experimental installation (see Fig.~2): the initial discharge
voltage $U_{\mathrm{dis}.}= $ $=3.8~\mathrm{kV}$
($Q\approx4~\mathrm{kJ}$, $P\approx4$~\textrm{MW}); the pressure of
the ignition gas, Ar, $p=0.8$~Pa
($N_{0}\approx2\times10^{14}$~\textrm{cm}$^{-3}$); the magnitude of
magnetic field induction was set by establishing three values of
voltage $U_{B}$ at the capacitor bank of the magnetic system,
$U_{B}=0.85$, 1.4, and 2$~\mathrm{kV}$ (see Fig.~3).

Under those initial conditions, a gas-metal plasma with the density
$N_{p}\sim10^{14}$~\textrm{cm}$^{-3}$ emerges in the discharge. This
means that the plasma layers exist with $N_{\rm
cr}^{1}=1.7\times10^{13}$~\textrm{cm}$^{-3}$ and $N_{\rm
cr}^{2}=6.5\times10^{13}$~\textrm{cm}$^{-3}$, which allowed us to
use the sounding frequencies $f^{1}=$ $=37.13~\mathrm{GHz}$ and
$f^{2}=72.88~\mathrm{GHz}$. The times of the existence of plasma
layers with $N_{\rm cr}^{1}=1.7\times10^{13}$~\textrm{cm}$^{-3}$ and
$N_{\rm cr}^{2}=6.5\times10^{13}$~\textrm{cm}$^{-3},$ which were
determined with the help of an UHF interferometer~-- i.e. the times
of the microwave signal cut-off $\tau_{\mathrm{cutoff}}$~-- equaled
about 1.8 and 1~ms, respectively, at $U_{B}=1.4~\mathrm{kV}$
\cite{17}. Typical oscillograms of signals obtained from the UHF
interferometer and UHF reflectometers are shown in Fig.~5. The
behavior of ACF and CCF for the reflected signal is illustrated in
Fig.~6.

The dynamics of a reflex-discharge plasma can be conditionally divided
into a few stages \cite{18,19}. At the first stage, the plasma layers
with $N_{p}=N_{\rm cr.}^{1,2}$ and the radius equal to the sounding
wavelength, $r=\lambda$, emerge. The second stage includes the
presence of plasma layers with $N_{p}=N_{\rm cr.}^{1,2}$. At this
stage, the radial dimensions of plasma layers grow in time. In other
words, the plasma layers move toward the combined antennas of
reflectometer (see Fig.~4), until their radius reaches a certain
value $r=r_{\mathrm{max}}$. Then, the growth of plasma layers stops,
and their radii practically do not change during the time
interval $\Delta t$ (hundreds of microseconds). Further, the radial
dimensions of plasma layers diminish, and the third stage begins,
when the density decreases and the plasma decays.

The character of the phase variation for two microwave signals (see
Fig.~5) reflected from a plasma layer with uniform density and
poloidally separated at two points (see Fig.~4) testifies that the
phase shifts in different cross-sections at the same frequency are
identical at a fixed time moment, which confirms the symmetry of a
plasma column with respect to its axis. The maximum radius of plasma
layers with the critical density $N_{p}=N_{\rm cr.}^{1,2}$ grows
with the initial magnetic field and, at $B=0.2$~T
($U_{B}=2~\mathrm{kV}$), reaches the values
$r_{\mathrm{max}}^{1}\approx5$~$\mathrm{cm}$ and
$r_{\mathrm{max}}^{2}\approx3.9$~$\mathrm{cm}$ for $N_{\rm
cr}^{1}=1.7\times10^{13}$~\textrm{cm}$^{-3}$ and $N_{\rm
cr}^{2}=6.5\times10^{13}$~\textrm{cm}$^{-3}$, respectively.

The time dependences of the rotational velocities of plasma layers
with $N_{p}=1.7\times10^{13}$~\textrm{cm}$^{-3}$ (layer~${\bf A}$)
and $N_{p}=6.5\times10^{13}$~\textrm{cm}$^{-3}$ (layer~${\bf B}$)
obtained for a number of discharge pulses are plotted in Fig.~7.
Similar dependences were obtained for other values of magnetic
induction as well. One can see that the rotational velocity grows to
a certain value and then starts to fall down, with the maximum
velocity being observed at the radius $r=r_{\mathrm{max}}$. The
angular rotational velocities $\omega_{\varphi}$ for layers ${\bf
A}$ and ${\bf B}$ with different radii are also different, i.e.
$\omega_{\varphi}^{\mathrm{A}}\neq\omega_{\varphi}^{\mathrm{B}}$.

In Fig.~8, the dependences of the maximum rotational velocity
$v_{\mathrm{max}}$ on the magnitude of initial magnetic field $B$
are depicted. The dependences are increasing functions, i.e. the
values of $B$ and $v_{\mathrm{max}}$ grow together. When the
velocity increases, the radii of layers with $N_{p}=N_{\rm
cr.}^{1,2}$ also increase, which was mentioned above. In this case,
it is natural to assume that the velocity increases with the radius,
because they are coupled by the relation
$v_{\varphi}=\omega_{\varphi }r$ and may be independent, probably,
of the magnetic field. However, as the radius increases, the growth
of the angular velocity with $B$ is also experimentally observed,
which confirms, in essence, the growth of $v_{\mathrm{max}}$ with
$B$.\vspace*{-2mm}

\begin{figure}
\vskip1mm
\includegraphics[width=7cm]{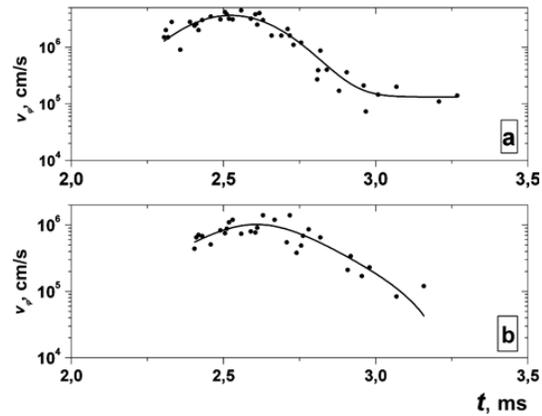}
\vskip-3mm\caption{Time dependences of the rotational velocity for
plasma layers with $N_{p}=1.7\times10^{13}$ (\textit{a}) and
$6.5\times10^{13}$~\textrm{cm}$^{-3} $\ (\textit{b}) for a mixture
Ar~+~Ti ($p=0.8$~Pa, $U_{\mathrm{dis}}=3.8~\mathrm{kV}$, and
$U_{\mathrm{B}}=2~\mathrm{kV}$)  }\vskip3mm
\end{figure}

\begin{figure}
\includegraphics[width=7cm]{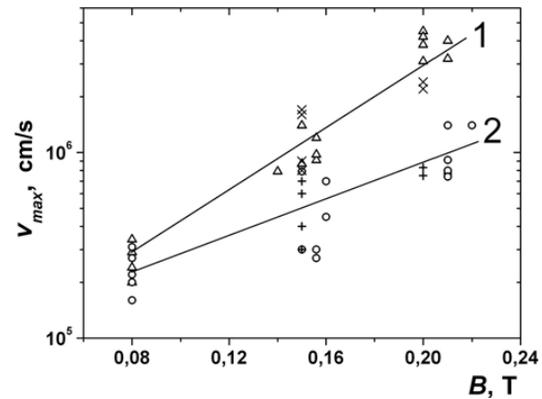}
\vskip-3mm\caption{Dependences of the maximum rotational velocity of
plasma (Ar + Ti) layers with $N_{p}=1.7\times10^{13}$
(\textit{1},$\bigtriangleup,\times$) and
$6.5\times10^{13}$~\textrm{cm}$^{-3}$\ (\textit{2},$\circ,+$) on the
magnetic field induction. $p=$ $=0.8$~Pa,
$U_{\mathrm{dis}}=3.8~\mathrm{kV}$, $\bigtriangleup$ and $\circ$ are
experimental data, $\times$ and $+$ correspond to calculation
results  }
\end{figure}

\section{Discussion of Experimental Results}

Experimentally, the difference between the angular velocities of
rotation for layers with $N_{p}=N_{\rm cr.}^{1,2}$ was observed
($\omega_{\varphi}^{\mathbf A} \neq \omega_{\varphi}^{\mathbf B} $).
If the plasma layers rotate as a whole (similarly to a solid
body), the angular velocity $\omega_{\varphi}$ of rotation for
layers with different radii would have been identical, and the
linear velocity would have grown proportionally to the radius. At
the same time, the results obtained testify that the model of plasma
rotation as a whole is inapplicable in this case. Similar
conclusions were made in work \cite{14} devoted to studying a weakly
ionized plasma of reflex discharge with the heated cathode.

The increase of the maximum rotational velocity $v_{\max}$ with the
growth of $B$ (see Fig.~8) can be qualitatively described in the
framework of the one-fluid MHD plasma model. As is seen from
Eq.~(\ref{15}), the radial component $j_{r}$ of the current density,
when interacting with the longitudinal magnetic field $B_{z}$,
stimulates the plasma rotation. In the first approximation, the
rotational velocity $v_{\varphi}$ is proportional to $j_{r}B_{z}$.
Then it follows that the rotational velocity also increases with the
magnetic field. The rotational velocity can be evaluated from\,
the\, relation  \mbox{$v_{\varphi}=-BQ(t)/2\pi r\rho$}, where the
quantity $Q(t)=$ $=\int_{0}^{t}2\pi rj_{r}dt$ was obtained in work
\cite{26} within the one-fluid MHD model. The results of our
calculations are shown in Fig.~8. One can see that they are in
satisfactory agreement with experimental results.

On the basis of experimental data, the magnitude of radial electric
field in plasma can be evaluated using formula (\ref{5}). The
calculations give the $E_{r}$-values in the intervals of 1.6--90 and
1.3--28~V/cm for layers ${\bf A}$ and ${\bf B}$, respectively, in
the examined range of magnetic fields. Making allowance for the
pressure gradient (formula (\ref{12})) gives a value for the radial
electric field that is slightly lower. As was mentioned above, the
rotational velocities of electrons and ions have to be different.
Knowing the electric field, one can evaluate the rotational velocity
for the ionic component by formula (\ref{4}). The values calculated
for Ar and Ti ions at $B=0.08\div0.2$~T lie in the interval
$(1.1\div9)\times10^{5}~\mathrm{cm/s}$ for both plasma layers (at
$B=0.2$~T, the rotational velocities in layer ${\bf A}$ equal
$(1.7\div2)\times10^{6}~\mathrm{cm/s}$). Those values of rotational
velocity for the ionic component are a bit lower or close to the
critical velocity $v_{\rm cr}=\left( 2e\phi _{i}/m_{i}\right)
^{1/2}$, where $\phi_{i}$ is the atomic ionization potential
\cite{27}. For argon, this quantity equals $v_{\rm cr}=8.7\times
10^{5}~\mathrm{cm/s}$. On the other hand, the examined plasma has
two basic ionic components (Ar and Ti), the critical velocities of
which are different. For instance, $v_{\rm
cr}=5.2\times10^{5}~\mathrm{cm/s}$ for Ti. According to the results
of work \cite{28}, the critical velocity for a two-component mixture
can be determined from the relation $ v_{\rm cr}=\sqrt{\frac{2\left(
\alpha e\phi_{i1}+\left( 1-\alpha\right) e\phi_{i2}\right) }{\alpha
m_{i1}+\left( 1-\alpha\right) m_{i2}}}, $ where $\phi_{i1}$ and
$\phi_{i2}$ are the corresponding ionization potentials for atoms in
the mixture, $m_{i1}$ and $m_{i1}$ are their masses, and $\alpha$ is
the content of component~1 in the mixture. The calculated value is
in satisfactory agreement with the experimental values of $v_{\rm
cr}$ obtained for various mixtures \cite{28}. The evaluation for the
two-component mixture of neutral Ar and Ti atoms with their content
of 1:1 gives the value $v_{\rm cr}=$ $=7\times10^{5}~\mathrm{cm/s}$.
This estimate is quite correct in the case where the mixture with
the given content is in the discharge volume before the discharge is
initiated (before the plasma formation).

According to the conditions of the described experiment, Ti atoms
are introduced in the direction along the column of the previously
created Ar plasma (see Fig.~2), where they become ionized owing to
their collisions with electrons (for Ti, the ionization potential is
lower and the ionization rates are higher than those for Ar), ions
(non-resonant charge exchange), and excited Ar atoms (Penning
ionization). Then, it follows that the content of neutral titanium
atoms in the main plasma column is insignificant. At 90\%~Ar +
10\%~Ti, the value $v_{\rm cr}=8.4\times10^{5}~\mathrm{cm/s}$, which
is close to $v_{\rm cr}$ for argon. On the other hand, we may
suppose that a high content of titanium atoms is characteristic of
the regions near the cathodes. In this case, the law of isorotation
(Ferraro's theorem) has to be taken into account, which reads that,
for a stationary motion, the angular velocity is constant along the
magnetic field lines. If the magnetic field has mirror geometry and
the isorotation is present, the maximum velocity equals
$v_{\mathrm{max}}=v_{\rm cr}R^{1/2}$, where $R$ is the mirror ratio
\cite{7}. Taking this reasoning into account and adopting $v_{\rm
cr}=7\times10^{5}~\mathrm{cm/s}$, we obtain that
$v_{\mathrm{max}}\approx8\times10^{5}~\mathrm{cm/s}$, which is close
to the value of $v_{\rm cr}$ for argon. Hence, as follows from the
consideration given above and the estimations made in the general
case, the rotational velocities of the ionic component are lower or
close to the critical value for argon.

\section{Conclusions}

1.~A specific feature of this work is the application of
two-frequency microwave fluctuation reflectometry for the
determination of the rotational velocity of plasma layers with
different densities.

2.~The difference between the angular rotational velocities of
plasma layers with different densities
($\omega_{\varphi}^{\mathbf{A}}\neq \omega_{\varphi}^{\mathbf{B}}$)
has been detected experimentally. The time dependences of rotational
velocities for plasma layers with different densities are
determined. The maximum of rotational velocity is achieved in 0.5~ms
after the discharge has begun. The reflectometric measurements have
confirmed the axial symmetry of the plasma column.

3.~The rotational velocity of plasma layers is found to grow with
the magnetic field in the examined interval of its values. The
estimates obtained for the rotational velocity in the framework of
the one-fluid MHD model are in satisfactory agreement with
experimental results.

4.~On the basis of experimental data, the radial electric field
strength in the plasma layers with various densities is evaluated.
We obtain $E_{r}=90\mathrm{~V/cm}$ at
$N_{p}=1.7\times10^{13}$~\textrm{cm}$^{-3}$ and
$E_{r}=28\mathrm{~V/cm}$ at
$N_{p}=6.5\times10^{13}$~\textrm{cm}$^{-3}$.

\vskip-2mm
\rezume{%
Ю.В. Ковтун, Є.І. Скібенко,\\ А.І. Скибенко, В.Б.
Юферов}{ДОСЛІДЖЕННЯ ОБЕРТАННЯ\\ ПЛАЗМОВИХ ШАРІВ РІЗНОЇ ГУСТИНИ\\ В
СХРЕЩЕННИХ $\mathbf{E}\times\mathbf{B}$ ПОЛЯХ} {Робота присвячена
визначенню швидкості обертання плазмових шарів різної густини у
плазмі імпульсного відбивного розряду. Для цього була запропонована
і використовувалась двочастотна НВЧ флуктуаційна рефлектометрія. За
її допомогою було встановлено відмінність кутових швидкостей
обертання плазмових шарів різної густини  та визначено їх часовий
хід. Також встановлено, що швидкість обертання плазмових шарів
зростає із збільшенням індукції магнітного поля. На основі одержаних
експериментальних даних оцінена величина напруженості радіального
електричного поля в плазмових шарах різної густини.}

\end{document}